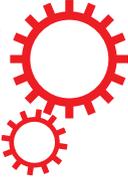

# Two-dimensional topological insulators with tunable band gaps: Single-layer HgTe and HgSe

Jin Li[1,2], Chaoyu He[1,2], Lijun Meng[1,2], Huaping Xiao[1,2], Chao Tang[1,2], Xiaolin Wei[1,2], Jinwoong Kim[3], Nicholas Kioussis[3], G. Malcolm Stocks[4] & Jianxin Zhong[1,2]

Two-dimensional (2D) topological insulators (TIs) with large band gaps are of great importance for the future applications of quantum spin Hall (QSH) effect. Employing *ab initio* electronic calculations we propose a novel type of 2D topological insulators, the monolayer (ML) low-buckled (LB) mercury telluride (HgTe) and mercury selenide (HgSe), with tunable band gap. We demonstrate that LB HgTe (HgSe) monolayers undergo a trivial insulator to topological insulator transition under in-plane tensile strain of 2.6% (3.1%) due to the combination of the strain and the spin orbital coupling (SOC) effects. Furthermore, the band gaps can be tuned up to large values (0.2 eV for HgTe and 0.05 eV for HgSe) by tensile strain, which far exceed those of current experimentally realized 2D quantum spin Hall insulators. Our results suggest a new type of material suitable for practical applications of 2D TI at room-temperature.

Topological insulators (TIs) have attracted extensive research interest due to the exotic properties and potential applications in spintronics and quantum computing[1–6]. 2D TIs are considered to be more promising materials than three-dimensional (3D) TIs for spin transport applications because of the robustness of conducting edge states from backscattering[7–9]. However, to date the quantum spin Hall (QSH) effect of existing 2D TIs occurs only at very low-temperatures, below 10 K for HgTe/CdTe[7] and InAs/GaSb[10] quantum wells and below $10^{-2}$ K for graphene[11,12], due to the small band gaps. Therefore, the search for 2D TIs with large band gaps has intensified in recent years for both fundamental and practical interests[9,13–18].

Since the discovery of graphene[19], 2D materials have been attracting great interest due to the various applications in optoelectronics, spintronics, solar cells[20,21]. Most of 2D materials, such as hexagonal BN, metal chalcogenides, transition metal dichalcogenides, etc., have 3D layered parent counterparts exhibiting strong in-plane covalent bonding and weak out-of-plane van der Waals (vdW) interactions, thus facilitating isolation of single layers via mechanical or chemical exfoliation. Furthermore, there are several materials beyond 3D layered materials that have stable 2D structures and exhibit a wide range of interesting properties. For instance silicene has been theoretically predicted[22–26] as a buckled honeycomb arrangement of Si atoms and having an electronic dispersion resembling that of relativistic Dirac fermions. Recently, silicene has been successfully grown on Ag[27–30], Ir[31] and ZrB$_2$[32] substrates and is considered as one of the most promising materials for the next-generation electronic devices[25,30,33]. Interestingly, recent *ab initio* and phonon-mode calculations have shown that the 2D honeycomb structure of many group-IV elements and III-V binary compounds are stable in planar or buckled geometries[24,25]. Thus, the honeycomb lattice of these materials offers a unique playground for searching of novel QSH insulators with large band gap.

[1]Hunan Key Laboratory of Micro-Nano Energy Materials and Devices, Xiangtan University, Hunan 411105, P. R. China. [2]Laboratory for Quantum Engineering and Micro-Nano Energy Technology and School of Physics and Optoelectronics, Xiangtan University, Hunan 411105, P. R. China. [3]Department of Physics, California State University, Northridge, California 91330-8268, USA. [4]Materials Science & Technology Division, Oak Ridge National Laboratory, Oak Ridge, Tennessee 37831, USA. Correspondence and requests for materials should be addressed to J.L. (email: lijin@xtu.edu.cn) or J.Z. (email: jxzhong@xtu.edu.cn)

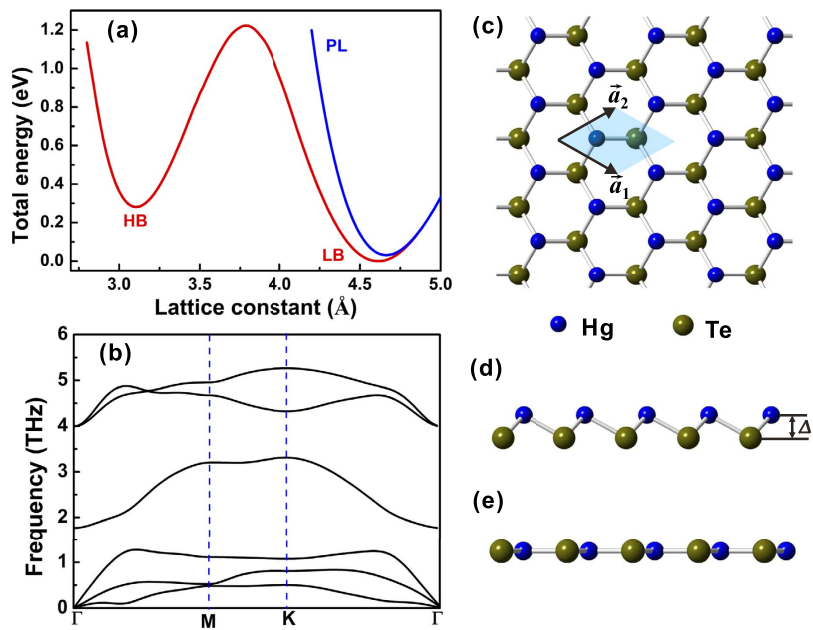

**Figure 1. Structural properties of monolayer HgTe.** (**a**) Total energies of 2D HgTe monolayer as a function of lattice constant for the planar (blue curve) and buckled (black curve) geometries, respectively. The low- and high-buckled geometries are denoted by LB and HB, respectively. (**b**) Phonon dispersion of the low-buckled honeycomb structure along high symmetry lines. (**c,d,e**) Top and side views of the 2D low-buckled and planar honeycomb structures, where the buckling distance Δ denotes the interlayer distance between Hg and Te planes. The unit cell is denoted by the shaded area.

The 3D HgSe and HgTe with the space group $F\bar{4}3m$ are semimetals having the unique property that their band structure exhibits an inversion of the $\Gamma_6$ and $\Gamma_8$ band ordering at the Brillouin zone (BZ) center[7,8,34,35], which is the origin of the QSH effect in 2D HgTe/CdTe quantum wells[7,8]. Consequently, HgSe and HgTe are also expected to have Dirac-like surface states which however are always coupled to metallic bulk states due to their semimetallic nature and hence difficult to be observed. However, applied strain induces a gap between the light- and heavy-hole bands rendering the systems 3D TIs[1,36,37]. In contrast to 2D TIs from group-IV elements, the mercury chalcogenides are composed of heavy elements having large spin orbit coupling (SOC) which combined with the 2D quantum confinement can lead to significant changes in the electronic band structure and the frontier band ordering compared with their 3D parent material. In the present work based on *ab initio* structure and phonon-mode calculations we predict that 2D monolayers of HgTe and HgSe are stable with low-buckled (LB) honeycomb geometry. By determining the uniaxial strain evolution of the band inversion in the band structure and of the $\mathbb{Z}_2$ topological invariant we demonstrate that LB HgTe undergoes a trivial insulator to topological insulator transition at $\varepsilon > 2.6\%$ uniaxial strain. Furthermore, the band gap of the TI phase can be tuned over a wide range from 0 eV to 0.2 eV as the tensile strain increases from 2.6% to 7.4%. In addition, we predict a similar strain-induced topological phase transition for the LB HgSe single layer at 3.1% strain, where the topological band gap increases to 0.05 eV at about 4.6%. The LB structure allows the formation of two different topological edge states in the zigzag and armchair edges.

## Results and Discussions

**HgTe monolayer.** Figure 1(a) displays the variation of the total energy as a function of the lattice constant of the 2D HgTe monolayer in planar (PL) and buckled (B) geometries shown in Fig. 1(c–e), respectively. One can see that the planar honeycomb structure has a single minimum, while the buckled structure has two minima corresponding to high-buckled (HL) and low-buckled (LB) honeycomb geometries separated by an energy barrier of about 0.3 eV. We find that the most stable structure is the LB with a lattice constant of 4.616 Å and buckling distance Δ = 0.47 Å. The phonon dispersion of the LB structure in Fig. 1(b) clearly shows that the acoustical branches and optical branches are well separated with a frequency gap. Furthermore, the absence of an imaginary frequency strongly suggests that the structure has a local minimum in its energy landscape. The stability of 2D monolayer HgTe under strain was corroborated by calculations of the phonon dispersion and the formation energies.

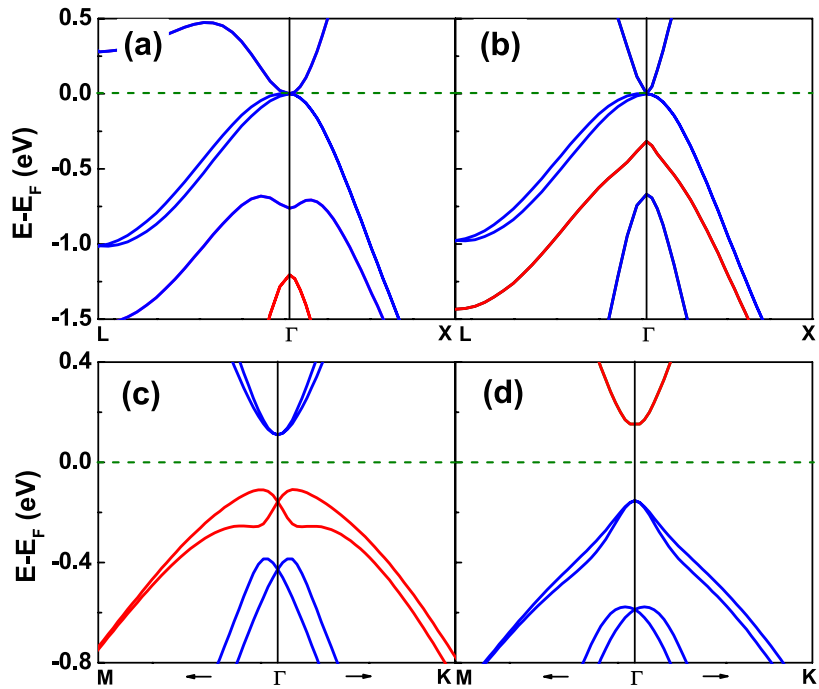

**Figure 2. Band structures of bulk and monolayer HgTe.** (**a**) Band structures of bulk HgTe employing the LDA and (**b**) MBJLDA exchange correlation functionals. Band structures of the 2D LB HgTe monolayer using the (**c**) LDA and (**d**) MBJLDA exchange correlation functionals. The Fermi level is denoted by the dashed lines at 0 eV. The red and blue lines denote the bands containing the $s$-like and $p$-like states at $\Gamma$ point, respectively.

The LDA and MBJLDA band structures of bulk HgTe along two symmetry directions are shown in Fig. 2(a,b), respectively. Both yield a semimetal which has an inverted band ordering ($p$-derived $\Gamma_8$ bands are above the $s$-derived $\Gamma_6$ band) where the heavy-hole and light-hole $\Gamma_8$ bands are degenerate at the BZ center. LDA places the $s$-type $\Gamma_6$-band 1.20 eV below the $\Gamma_8$-band and 0.44 eV below the spin-orbit split $\Gamma_7$-band, thus yielding the wrong band ordering between the $\Gamma_6$ and $\Gamma_7$ bands. On the other hand, the MBJLDA predicts the correct band ordering with $E_0 = E(\Gamma_6) - E(\Gamma_8) = 0.32$ eV and $\Delta_0 = E(\Gamma_8) - E(\Gamma_7) = 0.66$ eV, in good agreement with experiment and previous *ab initio* calculations[7,30,31]. The underlying origin is the poor treatment of the $p-d$ hybridization within LDA between the semicore Hg 5$d$ states and the Te-$p$ valence states, which in turn shifts the Te 5$p$ states to higher energies[38].

The LDA and MBJLDA band structures of the LB HgTe monolayer along two symmetry directions are shown in Fig. 2(c,d), respectively. Note, that the LDA, used commonly in the literature to determine the band topology, yields the incorrect ordering of the frontier bands at $\Gamma$ where the $s$-derived $\Gamma_4$ state lie above the $p$-derived $\Gamma_{5,6}$ states with an inverted band gap of 0.27 eV. Consequently, the LDA predicts incorrectly that the LB HgTe monolayer is a topological insulator with $\mathbb{Z}_2 = 1$. Similar artificial inverted bands have been also found by LDA calculations in other systems, such as ScAuPb and YPdBi, and can be corrected by MBJLDA[39]. In sharp contrast, the MBJLDA results predict that the $s$-derived $\Gamma_4$ states are about 0.31 eV higher than the $p$-derived $\Gamma_{5,6}$ states and $\mathbb{Z}_2 = 0$, indicating that the LB HgTe monolayer at equilibrium is a trivial insulator. These results demonstrate the necessity to use MBJLDA for accurate determination of the band topology.

In order to demonstrate the strain-induced band inversion process, we present in Fig. 3(a) the MBJLDA band energies of the $s$-derived $E(\Gamma_4)$, $p$-derived $E(\Gamma_{5,6})$, the band inversion strength, $E_I \equiv E(\Gamma_{5,6}) - E(\Gamma_4)$, and the absolute value of the global band gap $|E_g|$ as a function of in-plane tensile strain where the LB geometry is the ground state up to 7.4%. We find that the even though both energies $E(\Gamma_4)$ and $E(\Gamma_{5,6})$ decrease with tensile strain, the strain-induced change of the former is larger than that of the latter. Thus, we predict that the 2D HgTe honeycomb monolayer is a trivial insulator ($E_I < 0$) for $0 < \varepsilon < 2.6\%$ and undergoes a transition to a topological phase at 2.6% strain where $E_I$ vanishes and hence the band gap ($E_g = |E_I|$) closes. For $2.6\% < \varepsilon < 7.4\%$ the band ordering between the frontier $s$-derived $\Gamma_4$ and $p$-derived $\Gamma_{5,6}$ bands is inverted ($E_I > 0$) and $E_I$ increases with strain. It is important to emphasize that for $\varepsilon > 2.6\%$ $E_g < |E_I|$ due to the fact that the $s$-derived $\Gamma_4$ valence band develops a camel-back-shape (Fig. 4(c)) where the valence band maximum occurs in the vicinity of the BZ center. The topologically nontrivial band gap $E_g$ increases from 0 eV to 0.20 eV for $2.6\% < \varepsilon < 6.0\%$ and decreases to 0.14 eV for $6\% < \varepsilon < 7.4\%$ due to the enhancement of the camel-back-shape of the valence band. The large intrinsic band gap values of 0.0 ~ 0.2 eV in a wide range of tensile strain of 2.6% ~ 7.0% makes the 2D LB HgTe monolayer viable for room temperature applications.

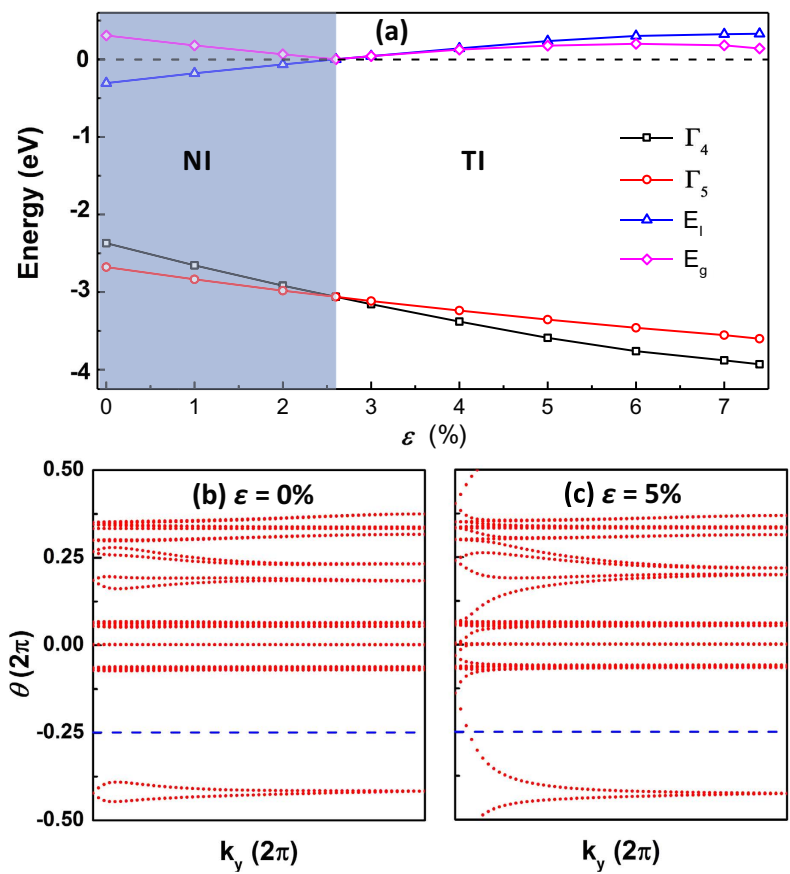

**Figure 3. Evolutions of the band structures and topological invariant $\mathbb{Z}_2$ of monolayer HgTe.** (**a**) Variation of the energies of the *s*-like $\Gamma_4$, *p*-like $\Gamma_{5,6}$, $E_I \equiv E(\Gamma_{5,6}) - E(\Gamma_4)$ and band gap $E_g$ as a function of tensile strain. The shaded and white areas denote the trivial and nontrivial topological phases, respectively. (**b,c**) Evolutions of Wannier function centers along $k_y$ for 2D HgTe monolayer under (**c**) 0.0% and (**d**) 5.0% tensile strain, respectively. The evolution curves cross any arbitrary reference line parallel to $k_y$ (for example, the blue dashed line) an odd (even) number of times, yielding $\mathbb{Z}_2 = 1\ (0)$.

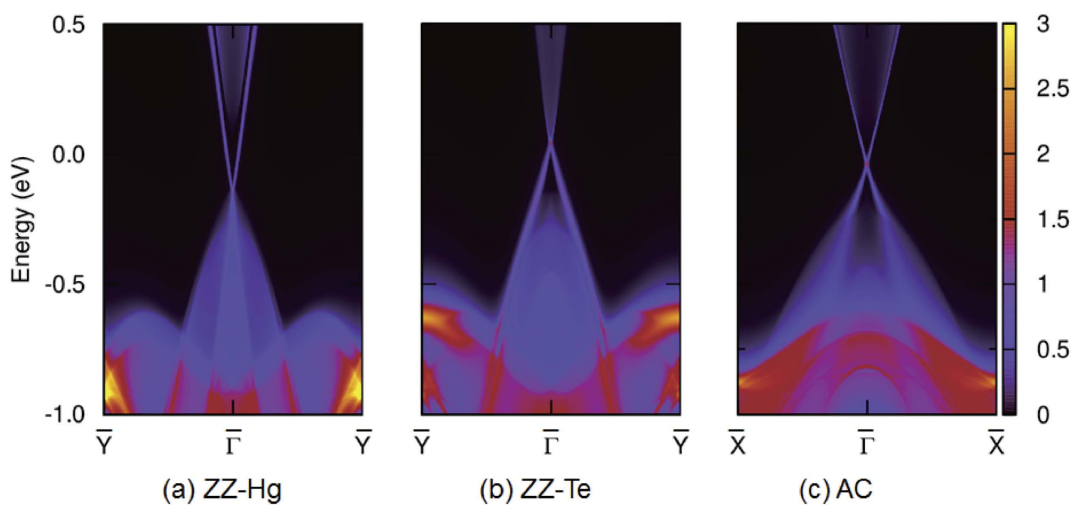

**Figure 4.** Edge states of LB 2D HgTe in the topological phase for (**a**) Hg- and (**b**) Te-terminated zigzag edges, respectively; and for (**c**) armchair edges.

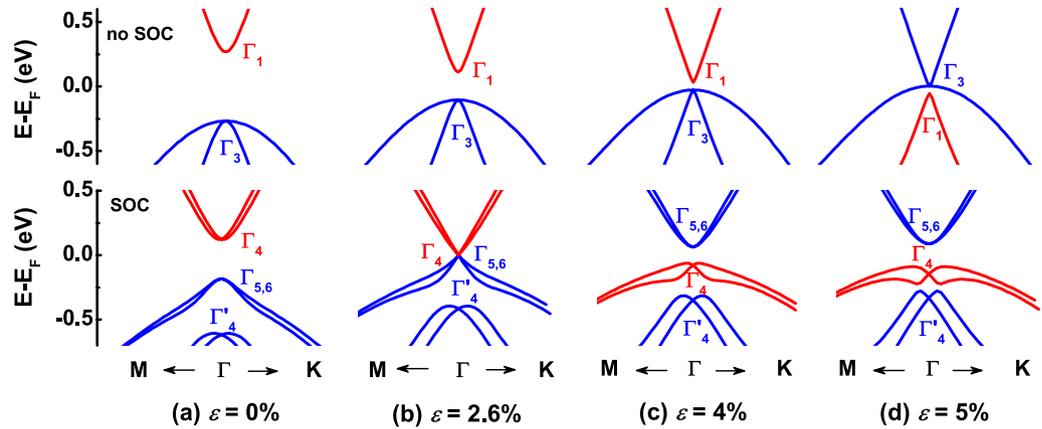

**Figure 5. Evolutions of the frontier bands of monolayer HgTe.** (**a**–**d**) are the band structures of the frontier bands of 2D LB HgTe under in-plane tensile strain of $\varepsilon = 0\%$, 2.6%, 4% and 5%, respectively. The band structures without and with SOC are shown in the top and bottom panels, respectively. The red and blue bands denote the $s$- and $p$-derived states at $\Gamma$.

The topological phase transition can be further confirmed by calculating the topological invariant $\mathbb{Z}_2$ before and after the band inversion. Figure 3(b,c) show the evolution lines of the Wannier function centers along $k_y$ for LB HgTe monolayer under in-plane tensile strain of $\varepsilon = 0\%$ and $\varepsilon = 5\%$, respectively. The system is TI (trivial insulator) if the evolution curves of the Wannier function centers cross an arbitrary reference line parallel to $k_y$ (blue dashed line) an odd (even) number of times yielding $\mathbb{Z}_2 = 1$ (0). We find that under zero (5%) strain the evolution curves (blue) cross any arbitrary line parallel to the horizontal axis (for example, the red dotted line) an even (odd) number of points, thus yielding $\mathbb{Z}_2 = 1$, verifying the TI phase under strain higher than 2.6%. An important feature of topological insulators is the topologically protected gapless edge states. Therefore, we have calculated the band structure of the semi-infinite ribbon employing the Green's function method with tight-binding parameters from VASP and wannier90. In Fig. 4 we show the band structure along the 2D symmetry lines of the (a) Hg-terminated zigzag (ZZ-Hg) edges, (b) Te-terminated zigzag (ZZ-Te) edges, and (c) armchair (AC) edges. Since the zigzag ribbon has an electric polarization, the Hg (Te) side Dirac-point is shifted towards the valence (conduction) band. The intrinsic misalignment of Dirac-points at opposite edges implies that the LB 2D HgTe has potential in spintronic applications since the dominant current channel among two (Hg and Te) edges can be easily tuned by the chemical potential.

Figure 5 shows the change of the frontier bands of the LB HgTe monolayer near the BZ center under in-plane tensile strain without and with SOC, respectively. Under zero strain and in the absence of SOC, the $s$-derived $\Gamma_1$ conduction band is about 0.54 eV above the four-fold degenerate $p$-derived $\Gamma_3$ valence band. In the presence of SOC, the $\Gamma_1$ band becomes $\Gamma_4$ and the $\Gamma_3$ band splits into two double-degenerate bands, the $p$-derived $\Gamma_{5,6}$ and $\Gamma'_4$ bands. The SOC energy, $E_{soc} = E(\Gamma_{5,6}) - E(\Gamma'_4)$, is about 0.43 eV which in turn reduces the band gap to 0.31 eV. Thus, there is no band inversion and the 2D HgTe monolayer is topologically trivial. At 2.6% strain and in the absence of SOC, the anti-bonding $\Gamma_1$ state shifts down with respect to the bonding $\Gamma_3$ state and hence the band gap deceases to 0.22 eV compared to its 0.54 eV value at zero strain. The SOC shifts the valence $\Gamma_{5,6}$ band up in energy resulting in the touching of the valence and $\Gamma_4$ conduction band at the Fermi energy. Note that in the absence of SOC the gap is not closed for 2.6% < $\varepsilon$ < 4.2% [Fig. 5(c)], while the system has already undergone a transition to a TI phase in the presence of SOC. This demonstrates that the SOC is the main driving force for the band inversion in the range of 2.6% < $\varepsilon$ < 4.2% strain. In sharp contrast, for $\varepsilon$ > 4.2% the band inversion shown in Fig. 5(d) already takes place *without* spin-orbit coupling but the system is semimetal due to the fact that the heavy- and light-hole $p$-derived $\Gamma_3$ bands are degenerate at the BZ center. SOC then opens up the gap for $\varepsilon$ > 4.2% rendering the LB HgTe a TI. Therefore, the strain and SOC have synergistic effects: In the range of 2.6% < $\varepsilon$ < 4.2%, strain reduces the band gap so that band inversion can be induced by SOC, while for $\varepsilon$ > 4.2% the SOC reopens the band gap of the already topologically inverted bands induced by strain.

To illustrate the origin of the strain-induced topological transition near $\Gamma$, we show schematically in Fig. 6(a–c) the evolution of the energy levels of the atomic levels under in-plane tensile strain including the effects of chemical bonding and SOC. The states near Fermi level are mainly Hg $s$-derived and Te $p$-derived orbitals which hybridize and split into the bonding and antibonding states, denoted by $s^+$ ($p^+$) and $s^-$ ($p^-$) with different parities. The SOC splits the $p^+$ state into the $\left|p, \pm \frac{3}{2}\right\rangle$ and $\left|p, \pm \frac{1}{2}\right\rangle$ with a total angular momentum of $j=\frac{3}{2}$ and $j=\frac{1}{2}$, respectively. In the absence of strain, because of the strong hybridization of the Hg- and Te-derived orbitals, the energy of $s^-$ is much higher than that of the $p^+$ shown in Fig. 6(a) and the SOC splitting ($\Delta_{soc}$) of $p^+$ is insufficient to induce the band inversion. With increasing in-plane tensile strain, the He-Te interlayer bond length increases and hence the Hg- and Te-derived orbitals hybridization decreases leading to a reduction of the splitting between the bonding and antibonding states. Under tensile strain in the range of 2.6% ~ 4.2%, the energy difference between $s^-$ and $p^+$ decreases but the energy of $s^-$ is still higher than that of the $p^+$ without SOC. Turning on the SOC, the energy of $|p_{\pm\frac{3}{2}}\rangle$ states becomes higher than that of $s^-$ inducing the band inversion and the phase transition to the topologically nontrivial phase as shown in Fig. 6(b). Further increasing the tensile strain, the energy of $s^-$ becomes lower than that of $p^+$ and the band inversion is achieved solely by the strain. However, there is no band gap as the Fermi level crosses the degenerate $p^+$ states. By including SOC, the topological band gap is reopened by the splitting of $p^+$ into the $\left|p,\pm\frac{3}{2}\right\rangle$ and $\left|p,\pm\frac{1}{2}\right\rangle$ and the 2D monolayer HgTe becomes TI. The change of the band order by strain in 2D monolayer HgTe is consistent with the previous theoretical study on the strain tuning of topological band order in cubic semiconductors, as increasing the lattice constant leads to a decrease of the coupling potentials[40].

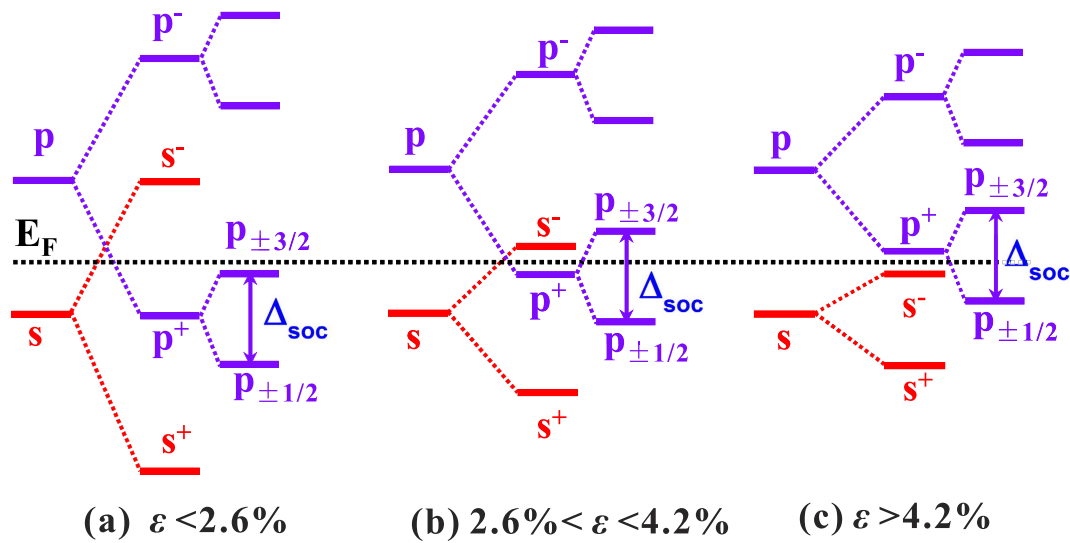

**Figure 6. Schematic diagram of topological transition.** The evolution of the energy levels at Γ point for the 2D HgTe monolayer with increasing in-plane tensile strain. The red and blue solid lines denote the *s*-derived and *p*-derived states. The superscripts "+" and "−" denote the bonding and antibonding states, respectively. The $|p^{\pm}\rangle$ states are split by the SOC ($\Delta_{soc}$) into the $|p_{\pm\frac{3}{2}}\rangle$ and $|p_{\pm\frac{1}{2}}\rangle$ state. The Fermi level is shown by the horizontal black dashed line.

We have also examined the origin of band splitting in Fig. 5(c) which can be either of Rashba[41] and/or Dresselhaus[42] spin-orbit coupling since both structural and bulk-inversion symmetries are broken. As was recently demonstrated by Fu[43], under $C_{3v}$ symmetry (which is the point group symmetry for HgTe) the Dresselhaus-type spin-orbit coupling vanishes along the Γ-M symmetry direction. Analysis of the **k**-resolved energy difference between the two top-most valence bands in Fig. 5(c) shows that further away from the Γ point the Dresselhaus-type spin-orbit coupling is dominant because the band splitting diminishes along the Γ-M point symmetry line. On the other hand, in the vicinity of Γ point the Rashba effect becomes dominant. We find that different orbital character of the valence bands (Hg-*s* and Te-*p*) induces distinct spin-orbit coupling effects.

**HgSe monolayer.** These results invite the intriguing question whether the 2D single layer of HgSe undergoes a similar TI transition under tensile strain. We find that the HgSe honeycomb monolayer has stable low-buckled structure with $a=4.35$ Å and $\Delta=0.368$ Å. The stability of 2D monolayer HgSe is confirmed by the phonon dispersion as shown in Fig. 7(e), even there are very small imaginary frequencies near Γ point, presumably these small imaginary frequencies will be removed using larger supercell sizes[26]. Figure 7 shows the MBJLDA band structure of the frontier bands of the LB HgSe monolayer under tensile strain $\varepsilon=0\%$, 3.1% and 4% without and with SOC, respectively. Under zero strain and without SOC, the *s*-derived $\Gamma_1$ band is about 0.42 eV higher than the *p*-derived $\Gamma_3$ band. Turning on SOC reduces the band gap between the *s*-like $\Gamma_4$ and *p*-like $\Gamma_{5,6}$ to 0.36 eV. As the tensile strain increases to 3.1%, the band gap without SOC decreases significantly to 0.07 eV. Inclusion of SOC closes the band gap with the $\Gamma_4$ conduction band touching the $\Gamma_{5,6}$ valence band. For strain greater than 3.3% the HgSe LB monolayer exhibits a band inversion [Fig. 7(c)] even in the absence of SOC. However, the SOC facilitates the reopening of the band gap. For $3.1\% < \varepsilon < 4.6\%$ the band gap of the topological phase increases to 0.05 eV, which is much larger than that for graphene and CdTe/HgTe/CdTe QW. For strain greater than 4.6% HgSe monolayer becomes planar and will be discussed elsewhere.

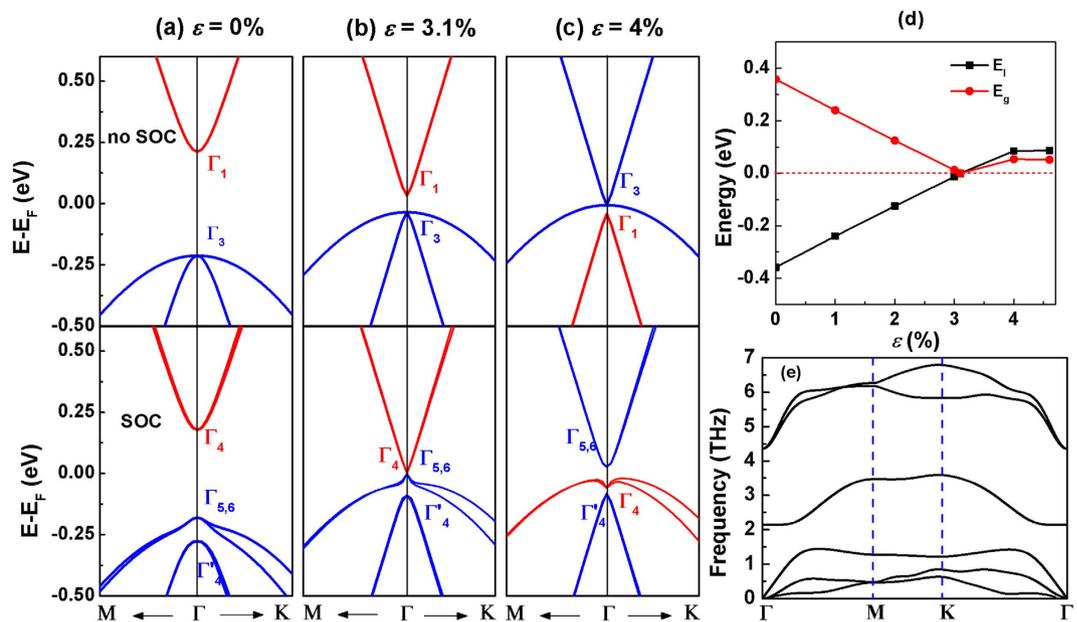

**Figure 7. Evolutions of the band structures of monolayer HgSe.** MBJLDA band structures of the frontier bands of the 2D LB HgSe under in-plane tensile strain of (**a**) $\varepsilon = 0\%$, (**b**) 3.1% and (**c**) 4%. The band structures without and with SOC are shown in the upper and bottom panels, respectively. The Fermi level is at 0 eV. The red and blue bands denote the $s$- and $p$-derived states at $\Gamma$, respectively. (**d**) The band inversion strength $E_I$ and band gap $E_g$ as a function of in-plane tensile strain. (**e**) The phonon dispersion of the low-buckled honeycomb HgSe at $\varepsilon = 0\%$.

Although HgTe and HgSe thin films have been experimentally studied extensively[8,44–46], the atomic-layer thick 2D HgTe and HgSe monolayers have not been reported so far. Based on the successful fabrication of the aforementioned 2D atomic-layer materials, a possible way to synthesize 2D-LB HgTe and HgSe monolayers is to grow them on appropriate substrates. For example, the honeycomb ZnO bilayer[47] and buckled silicene have been successfully synthesized on metal substrates such as Ag[27–30], Ir[31] and ZrB$_2$[32]. Other hybrid 2D superlattices have been successfully synthesized on 2d van der Waals substrates such as graphene/hexagonal boron nitride[48,49], and graphene/MoS$_2$[50–52]. Therefore, the 2D vdW nanostructures are also potential candidate substrates for growing of 2D LB HgTe and HgSe monolayers.

## Conclusion

In conclusion, based on *ab initio* structure and phonon-mode calculations we predict that the 2D LB HgTe and HgSe monolayers are stable. By examining the band inversion in the band structure and the evolution of the topological invariant, $\mathbb{Z}_2$, we demonstrate that the 2D LB HgTe monolayer undergoes a transition from a topologically trivial to a topological insulator phase at $\varepsilon > 2.4\%$ in-plane tensile strain. The underlying origin of the topological transition is the interplay of the SOC and strain. Most remarkably, the band gap of the 2D-LB HgTe TI phase can be tuned over a wide range from 0 eV to 0.2 eV as the tensile strain increases from 2.6% to 7.4%. Furthermore, we predict that the 2D LB HgSe monolayer underdoes a similar strain-induced topological insulator phase transition at $\varepsilon > 3.1\%$ where the topological insulator band gap is 0.05 eV at about 4.6%. The large band gap of the LB HgTe and HgSe monolayers make these 2D quantum spin Hall insulators suitable for practical applications at room-temperature.

## Methods

The calculations of the structural properties and the phonon dispersions were performed by using the Vienna Ab initio Simulation Package (VASP)[53] and phonopy[54] code, with the projector-augmented-wave[55] approach and the local density approximation[56]. An accurate description of the electronic structure is a prerequisite in the search and discovery efforts for the next-generation TIs. Therefore, the *ab initio* electronic structure calculations employed the full-potential WIEN2K code[57] with the SOC included and the modified Becke-Johnson Local Density Approximation (MBJLDA) functional[58]. The MBJLDA functional has been shown to yield accurate band gaps, effective masses, and frontier-band ordering at time-reversal

www.nature.com/scientificreports/invariant momenta (TRIM) that are in very good agreement with the computationally more intense GW and hybrid-functional approaches[59]. The $\mathbb{Z}_2$ topological invariant which characterizes the global band topology of the occupied Bloch wave functions in the entire BZ is determined by calculating the evolution of the Wannier function center (WFC) in reciprocal space during a "time-reversal pumping" process[37,60,61]. The edge state of a semi-infinite ribbon is calculated by using the Green's function method with the tight-binding parameters determined from wannier90[62].

## References

1. Fu, L. & Kane, C. L. Topological insulators with inversion symmetry. *Phys. Rev. B* **76,** 045302 (2007).
2. Hasan, M. Z. & Kane, C. L. Colloquium: Topological insulators. *Rev. Mod. Phys.* **82,** 3045 (2010).
3. Qi, X.-L. & Zhang, S.-C. The quantum spin Hall effect and topological insulators. *Phys. Today* **63,** 33 (2010).
4. Qi, X.-L. & Zhang, S.-C. Topological insulators and superconductors. *Rev. Mod. Phys.* **2010,** 83, 1057 (2010).
5. Peng, H. L. *et al.* Aharonov-Bohm interference in topological insulator nanoribbons. *Nature Materials* **9,** 225–229 (2010).
6. Kong, D. S. *et al.* Topological Insulator Nanowires and Nanoribbons. *Nano Lett.* **10,** 329–333 (2010).
7. Bernevig, B. A., Hughes, T. L. & Zhang, S.-C. Quantum Spin Hall Effect and Topological Phase Transition in HgTe Quantum Wells. *Science* **314,** 1757–1761 (2006).
8. König, M. *et al.* Quantum Spin Hall Insulator State in HgTe Quantum Wells. *Science* **318,** 766–770 (2007).
9. Chuang, F.-C. *et al.* Prediction of Large-Gap Two-Dimensional Topological Insulators Consisting of Bilayers of Group III Elements with Bi. *Nano Lett.* **14,** 2505–2508 (2014).
10. Knez, I., Du, R.-R. & Sullivan, G. Andreev Reflection of Helical Edge Modes in InAs/GaSb Quantum Spin Hall Insulator. *Phys. Rev. Lett.* **109,** 186603 (2012).
11. Min, H. K. *et al.* Josephson effect in ballistic graphene. *Phys. Rev. B* **74,** 041401 (2006).
12. Yao, Y. G., Ye, F., Qi, X.-L., Zhang, S.-C. & Fang, Z. Spin-orbit gap of graphene: First-principles calculations. *Phys. Rev. B* **75,** 041401(R) (2007).
13. Xu, Y. *et al.* Large-Gap Quantum Spin Hall Insulators in Tin Films. *Phys. Rev. Lett.* **111,** 136804 (2013).
14. Si, C. *et al.* Functionalized germanene as a prototype of large-gap two-dimensional topological insulators. *Phys. Rev. B* **89,** 115429 (2014).
15. Chou, B.-H. *et al.* Hydrogenated ultra-thin tin films predicted as twodimensional topological insulators. *New J. Phys.* **16,** 115008 (2014).
16. Wrasse, E. O. & Schmidt, T. M. Prediction of Two-Dimensional Topological Crystalline Insulator in PbSe Monolayer. *Nano Lett.* **14,** 5717 (2014).
17. Ma, Y. D., Dai, Y., Kou, L. Z., Frauenheim, T. & Heine T. Robust Two-Dimensional Topological Insulators in Methyl-Functionalized Bismuth, Antimony, and Lead Bilayer Films. *Nano Lett.* **15,** 1083 (2015).
18. Ma, Y. D., Dai, Y., Wei, W., Huang, B. B. & Whangbo, M.-H. Strain-induced quantum spin Hall effect in methyl-substituted germanane GeCH$_3$. *Sci. Rep.* **4,** 7297 (2014).
19. Novoselov, K. S. *et al.* Electric field effect in atomically thin carbon films. *Science* **306,** 666–669 (2004).
20. Xu, M. S., Liang, T., Shi, M. M. & Chen, H. Z. Graphene-Like Two-Dimensional Materials. *Chem. Rev.* **113,** 3766–3798 (2013).
21. Ivanovskii, A. L. Graphene-based and graphene-like materials. *Russ. Chem. Rev.* **81,** 571–605 (2012).
22. Takeda, K. & Shiraishi, K. Theoretical possibility of stage corrugation in Si and Ge analogs of graphite. *Phys. Rev. B* **50,** 14916 (1994).
23. Zhang, Y., Tan, Y.-W., Stromer, H. L. & Kim, P. Experimental observation of the quantum Hall effect and Berry's phase in graphene. *Nature* **438,** 201–204 (2005).
24. Durgun, E., Tongay, S. & Ciraci, S. Silicon and III-V compound nanotubes: Structural and electronic properties. *Phys. Rev. B* **72,** 075420 (2005).
25. Cahangirov, S., Topsakal, M., Aktürk, E., Şahin, H. & Ciraci, S. Two- and One-Dimensional Honeycomb Structures of Silicon and Germanium. *Phys. Rev. Lett.* **102,** 236804 (2009).
26. Şahin, H. *et al.* Monolayer honeycomb structures of group-IV elements and III-V binary compounds: First-principles calculations. *Phys. Rev. B* **80,** 155453 (2009).
27. Aufray, B. *et al.* Graphene-like silicon nanoribbons on Ag(110): A possible formation of silicene. *Appl. Phys. Lett.* **96,** 183102 (2010).
28. Lalmi, B. *et al.* & Aufray, B. Epitaxial growth of a silicene sheet. *Appl. Phys. Lett.* **97,** 223109 (2010).
29. Feng, B. J. *et al.* Evidence of Silicene in Honeycomb Structures of Silicon on Ag(111). *Nano Lett.* **12,** 3507–3511 (2012).
30. Sone, J., Yamagami, T., Aoki, Y., Nakatsuji, K. & Hirayama, H. Epitaxial growth of silicene on ultra-thin Ag(111) films. *New J. Phys.* **16,** 095004 (2014).
31. Meng, L. *et al.* Buckled Silicene Formation on Ir(111). *Nano Lett.* **13,** 685–690 (2013).
32. Fleurence, A. *et al.* Experimental Evidence for Epitaxial Silicene on Diboride Thin Films. *Phys. Rev. Lett.* **108,** 245501 (2012).
33. Liu, C. C., Feng, W. X. & Yao, Y. G. Quantum Spin Hall Effect in Silicene and Two-Dimensional Germanium. *Phys. Rev. Lett.* **107,** 076802 (2011).
34. Rohlfing, M. & Louie, S. G. Quasiparticle band structure of HgSe. *Phys. Rev. B* **57,** R9392 (1998).
35. Svane, A. *et al.* Quasiparticle band structures of β-HgS, HgSe, and HgTe. *Phys. Rev. B* **84,** 205205 (2011).
36. Brüne, C. *et al.* Quantum Hall Effect from the Topological Surface States of Strained Bulk HgTe. *Phys. Rev. Lett.* **106,** 126803 (2011).
37. Winterfeld, L. *et al.* Strain-induced topological insulator phase transition in HgSe. *Phys. Rev. B* **87,** 075143 (2013).
38. Wei, S. H. & Zunger, A. Role of metal *d* states in II-VI semiconductors, *Phys. Rev. B* **37,** 8958 (1998).
39. Feng, W. X., Xiao, D., Zhang, Y. & Yao, Y. G. Half-Heusler topological insulators: A first-principles study with the Tran-Blaha modified Becke-Johnson density functional. *Phys. Rev. B* **82,** 235121 (2010).
40. Feng, W. X. *et al.* Strain tuning of topological band order in cubic semiconductors. *Phys. Rev. B* **85,** 195114 (2012).
41. Bychkov, Y. A. & Rashba, E. I. Oscillatory effects and the magnetic susceptibility of carriers in inversion layers. *J. Phys. C* **17,** 6039 (1984).
42. Dresselhaus, G. Spin-orbit coupling effects in zinc blende structures. *Phys. Rev.* **100,** 580 (1955).
43. Fu, L. Hexagonal Warping Effects in the Surface States of the Topological Insulator Bi$_2$Te$_3$. *Phys. Rev. Lett.* **103,** 266801 (2009).
44. Rath, S., Paramanik, D., Sarangi, S. N., Varma, S. & Sahu, S. N. Surface characterization and electronic structure of HgTe nanocrystalline thin films. *Phys. Rev. B* **72,** 205410 (2005).
45. Kim, D.-W., Jang, J., Kim, H., Cho, K. & Kim, S. Electrical characteristics of HgTe nanocrystal-based thin film transistors fabricated on flexible plastic substrates. *Thin Solid Films* **516,** 7715–7719 (2008).
46. Hankare, P. P., Bhuse, V. M., Garadkar, K. M., Delekar, S. D. & Mulla, I. S. Chemical deposition of cubic CdSe and HgSe thin films and their characterization. *Semicond. Sci. Technol.* **19,** 70 (2004).
SCIENTIFIC REPORTS | 5:14115 | DOI: 10.1038/srep14115    8

## Acknowledgements

This work was supported by the National Natural Science Foundation of China (Grant No. 11404275, 11474244, 4040204, 11204261, 11304264, 11204260), the Program for Changjiang Scholars and Innovative Research Team in University (Grant No. IRT13093), the National Basic Research Program of China (Grant No. 2012CB921303, 2015CB921103), the Special Funds for Theoretical Physics of the National Natural Science Foundation of China (Grant No. 11347206), the Hunan Provincial Natural Science Foundation of China (Grant No. 13JJ4046) and the Specialized Research Fund for the Doctoral Program of Higher Education of China (Grant No. 20134301120004). The work at CSUN was supported by the NSF-PREM (Grant No. DMR-1205734). The work at ORNL was supported by the US Department of Energy, Office of Basic Energy Sciences.

## Author Contributions

J.L. conceived the work and performed most of the calculations, analyzed the data, drafted and revised the manuscript. C.Y.H. contributed to the calculations of phonon dispersions and participated in the discussions of the calculations and results. L.J.M., H.P.X., C.T., X.L.W. and G.M.S. participated in the discussions of the calculations and results. J.K. carried out the calculations of the edge states of semi-infinite ribbons and elucidated the underlying origin of the Rashba and Dresselhaus effects of the band structures. N.K. contributed to the Fig. 5, participated in the discussions of the calculations and results and revised the manuscript. J.X.Z. participated in the discussions of the calculations and results and revised the manuscript. All authors read and approved the final manuscript.

## Additional Information

**Competing financial interests:** The authors declare no competing financial interests.

**How to cite this article**: Li, J. *et al.* Two-dimensional topological insulators with tunable band gaps: Single-layer HgTe and HgSe. *Sci. Rep.* **5**, 14115; doi: 10.1038/srep14115 (2015).